\documentclass[sn-mathphys-num]{sn-jnl}

\usepackage{graphicx}%
\usepackage{multirow}%
\usepackage{amsmath,amssymb,amsfonts}%
\usepackage{amsthm}%
\usepackage{mathrsfs}%
\usepackage[title]{appendix}%
\usepackage{xcolor}%
\usepackage{textcomp}%
\usepackage{manyfoot}%
\usepackage{booktabs}%
\usepackage{algorithm}%
\usepackage{algorithmicx}%
\usepackage{algpseudocode}%
\usepackage{listings}%
\usepackage[version=4]{mhchem}%
\usepackage{lineno}

\theoremstyle{thmstyleone}%
\theoremstyle{thmstyletwo}%

\theoremstyle{thmstylethree}%

\begin{document}

\title{Anomalous refractive index modulation and giant birefringence in 2D ferrielectric \ce{CuInP2S6}}


\author*[1,3]{\fnm{Houssam} \sur{El Mrabet Haje}}\email{H.elMrabetHaje@tudelft.nl}
\equalcont{These authors contributed equally to this work.}

\author*[2,3]{\fnm{Roald J.H.}\spfx{van der} \sur{Kolk}}\email{R.J.H.vanderKolk@tudelft.nl}
\equalcont{These authors contributed equally to this work.}

\author[1,3]{\fnm{Trent M.}
\sur{Kyrk}}\email{t.m.k.kyrk@tudelft.nl}

\author[1,3]{\fnm{Mazhar N.} \sur{Ali}}\email{m.n.ali@tudelft.nl}

\affil[1]{\orgdiv{Department of Quantum Nanoscience, Faculty of Applied Sciences}, \orgname{Delft University of Technology}, \city{Delft}, \country{the Netherlands}}

\affil[3]{\orgdiv{Kavli Institute of Nanoscience}, \orgname{Delft University of Technology}, \city{Delft},  \country{the Netherlands}}

\affil[2]{\orgdiv{Department of Imaging Physics, Faculty of Applied Sciences}, \orgname{Delft University of Technology}, \city{Delft}, \country{the Netherlands}}


\abstract{\textbf{2D ferroelectric (FE) materials have opened new opportunities in non-volatile memories, computation and non-linear optics due to their robust polarization in the ultra-thin limit and inherent flexibility in device integration. Recently, interest has grown in the use of 2D FEs in electro-optics, demanding the exploration of their electronic and optical properties. In this work, we report the discovery of an unprecedented anomalous thickness-dependent change in refractive index, as large as $\delta n$ $\sim$ 23.2$\%$, in the 2D ferrielectric \ce{CuInP2S6}, far above the ultra-thin limit, and at room temperature. 
Furthermore, \ce{CuInP2S6} exhibits a giant birefringence in the blue-ultraviolet regime, with a maximum $\vert n_{OOP} - n_{IP}\vert$ $\sim$ 1.24 at $t \sim$ 22 nm and $\lambda$ = 339.5 nm, which is, to the best of our knowledge, the largest of any known material in this wavelength regime. We relate changes in \ce{CuInP2S6} optical constants to changes in the Cu(I) FE polarization contribution, influenced by its ionic mobility, opening the door to electronic control of its optical response for use in photonics and electro-optics.}} 

\maketitle

As a result of the increasing need for highly efficient, tunable, and high-speed electro-optic devices in modern-day telecommunications \cite{Sinatkas2021, wooten2000review}, sensing \cite{Hu2025} and quantum technologies \cite{arrazola2021quantum}, there has been a surge of interest in 2D FEs. They have shown non-linear and tunable optical and electronic properties \cite{Wang2023,Zhang2023,Mushtaq2022} with the potential to surpass contemporary 3D ferroelectric electro-optic platforms like \ce{LiNbO3} \cite{wooten2000review}. For instance, materials with large birefringence, particularly within the visible-ultraviolet range, are essential for light polarization manipulation \cite{Savage2009, Weber2000, Chen2018}, ultraviolet communications \cite{4511651}, and lithography \cite{Zhang2022xxx, Xu2022}, making them highly desired in the fields of photonics and integrated electro-optics. Typical platforms either suffer from low optical anisotropy \cite{Sinton1961, Ghosh1999, Ma2022}, high losses \cite{kats2012giant} or integration and stability concerns \cite{kats2012giant, zhang2011giant, Ma2022, wkeglowska2016high}. 2D FEs, due to their crystalline anisotropy and non-linear behaviors, present themselves as attractive new platforms in these fields. \\

Amongst the 2D FEs that have garnered significant attention, \ce{CuInP2S6} stands out due to its moderate band gap $E_g \sim$ 2.8 eV \cite{Zhou2021, Ho2021, Bu2023}, 
robust ferrielectricity at room temperature ($T_c$ $\sim$ 315 K) even to the ultra-thin limit \cite{Maisonneuve1997, Studenyak2003, Liu2016, Niu2019}, and remarkable ionic conductivity \cite{Zhu2023, Jiang2022}. This has led to observations of pyroelectricity down to the thin limit \cite{Niu2019}, giant negative piezoelectricity \cite{You2019}, an enhanced bulk photovoltaic effect \cite{Li2021}, and more \cite{Ming2022, Xue2025, Li2024, Si2019RoomTE}. 
Hence, CIPS is actively being investigated for its potential in varying fields, including nanoelectronics \cite{Io2022}, photonics \cite{Li2022}, electro-optics \cite{Shang2024}, and others \cite{Kong2022, Qiu2023, Jiang2024, Samulionis2001}.\\

The driving force behind the technologically relevant behaviors of CIPS are its Cu(I) cations, which contribute significantly to both its ferrielectricity and ionic conductivity \cite{Maisonneuve1997, Maisonneuvle1997, Xu2020, Zhang2021}. CIPS' crystal structure (Figure \ref{FIG1}a) is built from planar layers of filled sulfur octahedra, where the central voids are occupied by either P-P dimers, In(III) cations, or Cu(I) cations. At $T > T_c$, the Cu(I) ions are spatially disordered, giving rise to a paraelectric ionic conducting state. At room temperature, CIPS exists in a ferrielectric ionic conductive state due to a second-order Jahn-Teller distortion, where the Cu(I) cations partially occupy three positions: within the sulfur framework at a quasi-trigonal site, at an octahedral site, and within the van der Waals gap in a quasi-tetrahedral site. Consequently, In(III) cations displace in the opposite direction with respect to Cu(I) cations, generating a spontaneous and non-compensated out-of-plane (OOP) polarization \cite{Maisonneuve1997, Maisonneuvle1997}. \\

The Cu(I) cations in CIPS also displace within the plane, generating a thickness-dependent in-plane (IP) polarization \cite{Deng2020, Hu2024}.  Below critical thickness $t_c$ (reported to be at $t_c \sim$ 90 nm), a structural phase transition occurs, where the molecular layers of CIPS slide along the $\Vec{a}$ direction, changing the crystal symmetry from monoclinic \textit{Cc} to trigonal \textit{P}3$_1$\textit{c} (Figure \ref{FIG1}a), and causing the IP polarization to disappear while preserving the OOP polarization \cite{Deng2020}. Other thickness-dependent behaviors in CIPS have been observed, including an increase of the electrocaloric effect at $t \sim$ 169 nm \cite{Si2019RoomTE}, an abrupt decrease of the piezoelectric coefficient $d_{33}$ at $t<$ 40 nm \cite{Io2022}, and an enhancement of the bulk photovoltaic effect at $t<$ 80 nm \cite{Li2021}, among others \cite{Li2024, Kong2022, Bai2024, Song2023, Xue2025}. However, as yet, CIPS's optical properties and their anisotropies have not been thoroughly investigated as a function of thickness. Of particular interest are its refractive index $n$, extinction coefficient $\kappa$, and dielectric function $\hat{\varepsilon} = \varepsilon_1 + i\varepsilon_2$, as well as its birefringence $\Delta$n and dichroism $\Delta\kappa$. \\ 

Here we investigate the thickness and wavelength dependencies of CIPS's in-plane and out-of-plane optical constants, as well as its optical anisotropy, from the bulk to the thin limit using Variable Angle Spectroscopic Ellipsometry (VASE). We also characterize CIPS's vibrational modes using Raman spectroscopy. We observe an anomalous change in optical constants as a function of thickness as large as $\delta n$ $\sim$ 23.2$\%$ and within $t \in [22, 170)$ nm, which we attribute to changes in CIPS' polarization stemming from the Cu(I) component and, potentially, its ion mobility. Moreover, we also show that CIPS exhibits giant intrinsic birefringence $\vert\Delta n \vert$ = $\vert n_{OOP} - n_{IP}\vert$ for the entire range of wavelengths studied $\lambda$ $\in$ [210.6, 1688.3] nm, with a maximum $\vert\Delta n\vert$ $\sim$ 1.24 at $t \sim$ 22 nm and $\lambda$ = 339.5 nm, which is, to the best of our knowledge, the largest of any known material in the blue-ultraviolet regime. Both discoveries, the anomalous change of refractive index with thickness and the giant birefringence in CIPS, reveal thickness as a new tuning parameter and lay the groundwork for 2D ferroelectrics to be used as 
non-linear and tunable platforms for electro-optical control, even far from the ultra-thin limit and in a broad spectral range.

\section*{Results}\label{sec2}
\begin{figure}[H]
\centering
\includegraphics[width=1.00\textwidth, keepaspectratio]{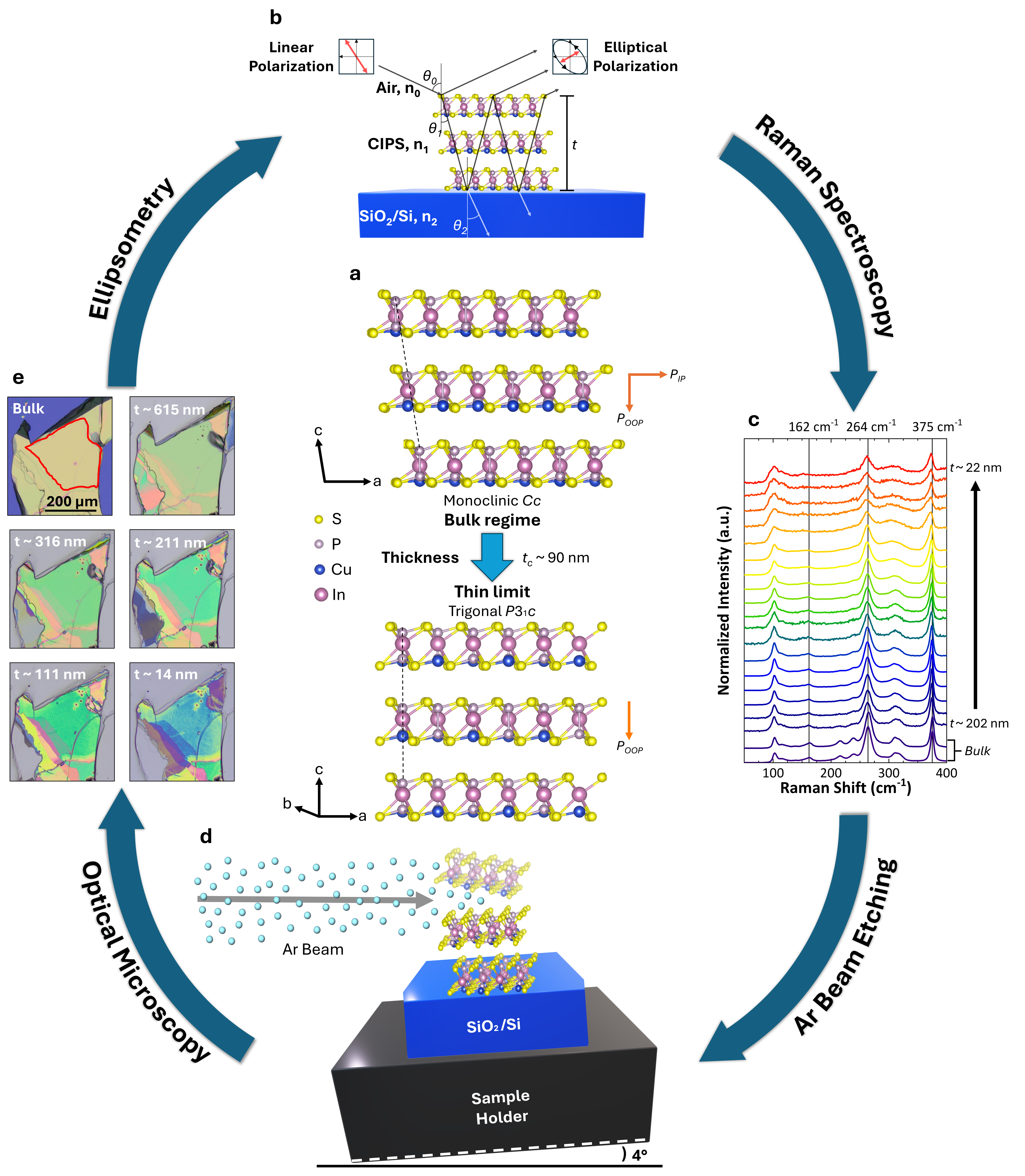}
\caption{\textbf{Experimental methodology cycle.} a) Side-view of the thickness-induced structural transition in CIPS from monoclinic \textit{Cc} \cite{cif} to trigonal \textit{P}3$_1$\textit{c} \cite{Deng2020}. Here, $t$ is the thickness of the CIPS flake and $t_{c}$ is the critical thickness at which the phase transformation occurs. The orange arrows indicate the in-plane ($\vec{P}_{IP}$) and out-of-plane ($\vec{P}_{OOP}$) polarizations. For $t > t_c$, the total polarization is $\vec{P}_T$ = $\vec{P}_{IP}$ + $\vec{P}_{OOP}$. For $t < t_c$, $\vec{P}_T$ = $\vec{P}_{OOP}$. b) Schematic of the spectroscopic ellipsometry measurements performed on CIPS, where $\theta_0$ is the incident angle, $\theta_1$ is the refracted angle of light in CIPS, and $\theta_2$ is the refracted angle of light in the substrate. The refractive index of light in air, CIPS, and SiO$_2$ is denoted as $n_0$, $n_1$, and $n_2$, respectively. c) Collection of all normalized CIPS' Raman spectra from bulk to $t \sim$ 22 nm. Black lines indicate peak positions of relevant vibrational modes.  d) Side-view schematic of the low-angle argon beam etching procedure used to reduce the thickness of the CIPS flake. e) Optical microscope images of the CIPS flake at different thicknesses, with the area studied by ellipsometry and Raman spectroscopy outlined in red.}
\label{FIG1}
\end{figure}

To investigate the thickness and wavelength-dependent optical properties of CIPS, a bulk flake  was exfoliated onto a SiO$_2$/Si substrate (see \nameref{sectionmethods} for details). 
The thickness and optical constants were extracted for wavelengths $\lambda \in$ [210.6, 1688.3] nm using VASE, as illustrated in Figure \ref{FIG1}b. 
Subsequently, Raman spectroscopy was performed to measure CIPS' vibrational modes  (Figure \ref{FIG1}c), followed by a thinning of the CIPS flake by approximately 10 nm using low angle argon beam etching to minimize damage to the flake's crystal structure (Figure \ref{FIG1}d). With this new thickness (Figure \ref{FIG1}e), the flake was again measured using VASE and Raman and the cycle was repeated until reaching a thickness of $t \sim$ 14 nm.  
Following this sequence, we were able to study the effects of thickness and wavelength on CIPS' optical constants and vibrational spectrum.

\subsection*{Effect of thickness on optical constants}




Figures \ref{n,k vs lambda FIG2} and \ref{Full n vs lambda} present CIPS's IP and OOP refractive indices and extinction coefficients at different thicknesses as a function of wavelength. 
CIPS's IP refractive index (Figure \ref{n,k vs lambda FIG2}a) peaks at  $\lambda \sim$ 350 nm. As thickness decreases, the maximum refractive index shifts to longer wavelengths but decreases in magnitude, from $n$ $\sim$ 3.6 at $t \sim$ 615 nm and $\lambda$ = 345.8 nm, to $n$ $\sim$ 3.5 at $t \sim$ 22 nm and $\lambda$ = 353.8 nm. However, there is a crossing point at $\lambda \sim$ 1320 nm (Figure \ref{Full n vs lambda}a), where for longer wavelengths the refractive index increases with decreasing thickness. Similarly, the maximum IP extinction coefficient (Figure \ref{n,k vs lambda FIG2}b) shifts to longer wavelengths as thickness decreases, but displays a crossing point at $\lambda \sim$ 274 nm, where the thickness dependence trend inverts. At $t \sim$ 615 nm and $\lambda \sim$ 456 nm, a kink is visible that continuously shifts to 
$\lambda \sim$ 441 nm at $t \sim$ 22 nm as thickness decreases, which aligns well with the band gap previously reported for CIPS of $E_g \sim$ 2.8 eV ($\lambda \sim$  443 nm) \cite{Zhou2021, Ho2021, Bu2023}. \\ 

In contrast, the OOP refractive index (Figure \ref{n,k vs lambda FIG2}c) and extinction coefficient (Figure  \ref{n,k vs lambda FIG2}d) exhibit completely different dependencies. For the OOP refractive index, the local maximum of $n$ $\sim$ 2.8 at $t \sim$ 615 nm and $\lambda =$ 280.5 nm continuously decreases in magnitude as the thickness decreases. At $t \in$ [60, 83) nm the feature becomes undetectable, and by $t \sim$ 22 nm it has merged with the minimum at shorter wavelengths already present in the bulk,  with $n$ $\sim$ 2.1 at $\lambda =$ 282.1 nm. Hence, around $\lambda \sim $ 280 nm, wavelength at which CIPS has been reported to be a high-performance ultraviolet photodetector \cite{Ma2020}, we obtain the largest change in refractive index with thickness. Simultaneously, a second peak emerges when $t \in$ (161, 170] nm, becoming a new local maximum with a refractive index of $n$ $\sim$ 2.4 at $t \sim$ 22 nm and $\lambda$ = 383.9 nm. 
Regarding the OOP extinction coefficient, there is a maximum in the extinction coefficient, with $\kappa$ $\sim$ 0.50 at $t \sim$ 615 nm and $\lambda =$ 269.5 nm, that decreases with thickness, flattening at $t \in $ [60, 83) nm. Concomitantly, a second peak emerges at $t \in$ (161, 170] nm, with a maximum in the OOP extinction coefficient of $\kappa$ $\sim$ 0.38 at $t \sim 22$ nm and $\lambda = $ 352.2 nm. The change in maxima in the OOP extinction coefficient results in a crossing point similar to that observed in the IP, but in this case at $\lambda =$ 295.0 nm.\\

When comparing the IP and OOP optical constants, CIPS unequivocally exhibits giant birefringence $\vert \Delta n \vert = \vert n_{OOP}  - n_{IP}\vert >$ 0.3 and dichroism $\vert \Delta \kappa \vert = \vert \kappa_{OOP} - \kappa_{IP} \vert >$ 0.3, both in the bulk regime, as previously demonstrated \cite{Dziaugys2013, Vysochanskii2003, Kohutych2022}), and most strongly in the few-nanometer regime (Figure \ref{Anisotropy vs thickness vs lambda}). The dichroism is locally maximized at $t \sim$ 22 nm and $\lambda$ = 253.5 nm, with $\vert \Delta \kappa \vert$ $\sim$ 1.55, while the birefringence is maximized at $t \sim$ 22 nm and $\lambda$ = 339.5 nm, with $\vert \Delta n \vert$ $\sim$ 1.24. 
$\Delta n $ changes sign as we decrease the wavelength from negative to positive at $\lambda \sim$ 244 nm, and remains giant for all measured thicknesses.\\


\begin{figure}[t]
\centering
\includegraphics[width=1.00\textwidth, keepaspectratio]{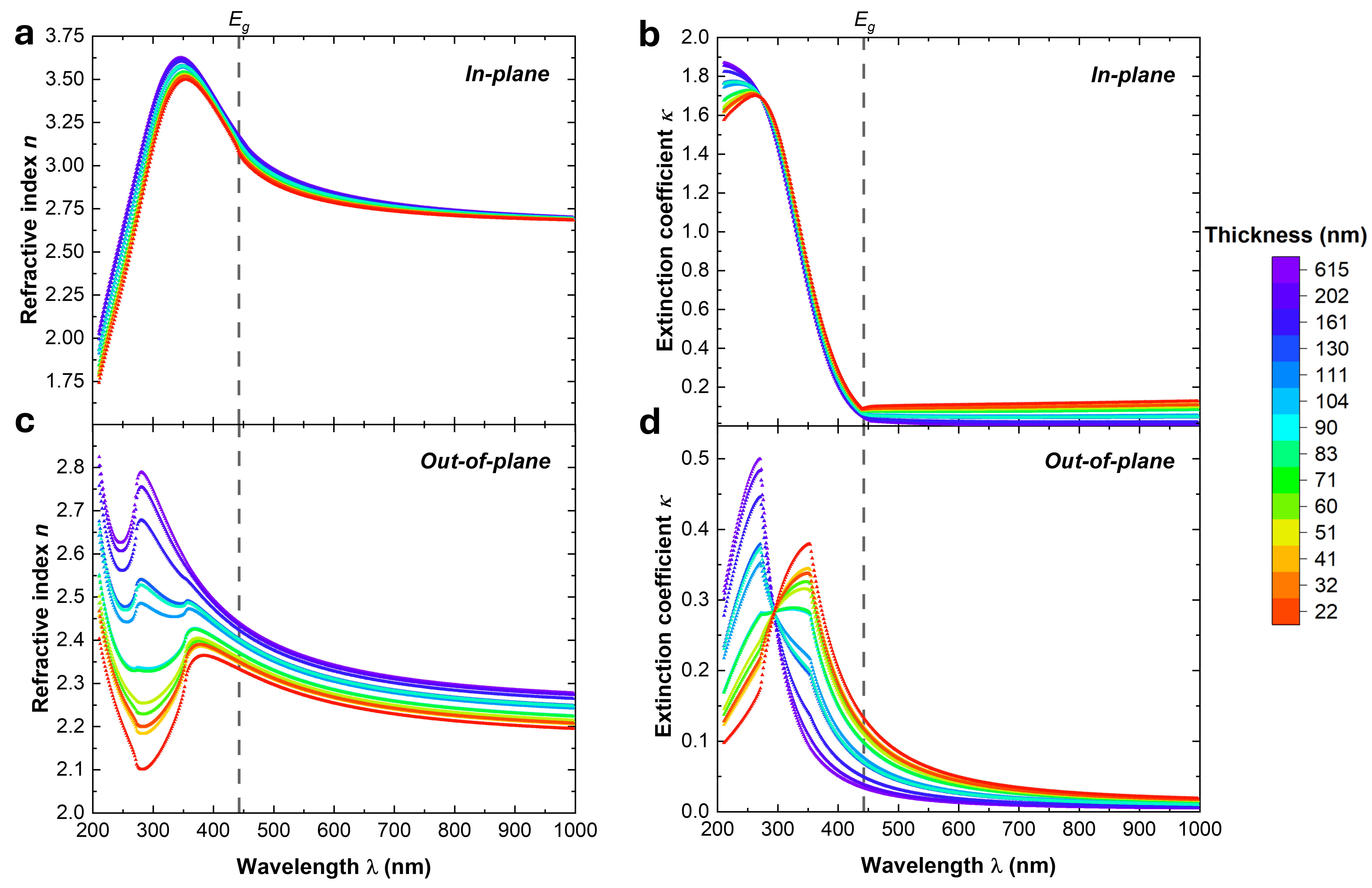}
\caption{\textbf{CIPS' optical constants wavelength dependence.} CIPS' wavelength-dependent  a) in-plane and c) out-of-plane refractive indices and b) in-plane and d) out-of-plane extinction coefficients at different thicknesses. The dashed lines correspond to the reported CIPS' band gap value $E_g \sim$ 2.8 eV ($\lambda \sim$  443 nm) \cite{Zhou2021, Ho2021, Bu2023}.}
\label{n,k vs lambda FIG2}
\end{figure}





\begin{figure}[t]
\centering
\includegraphics[width=1.0\textwidth, keepaspectratio]{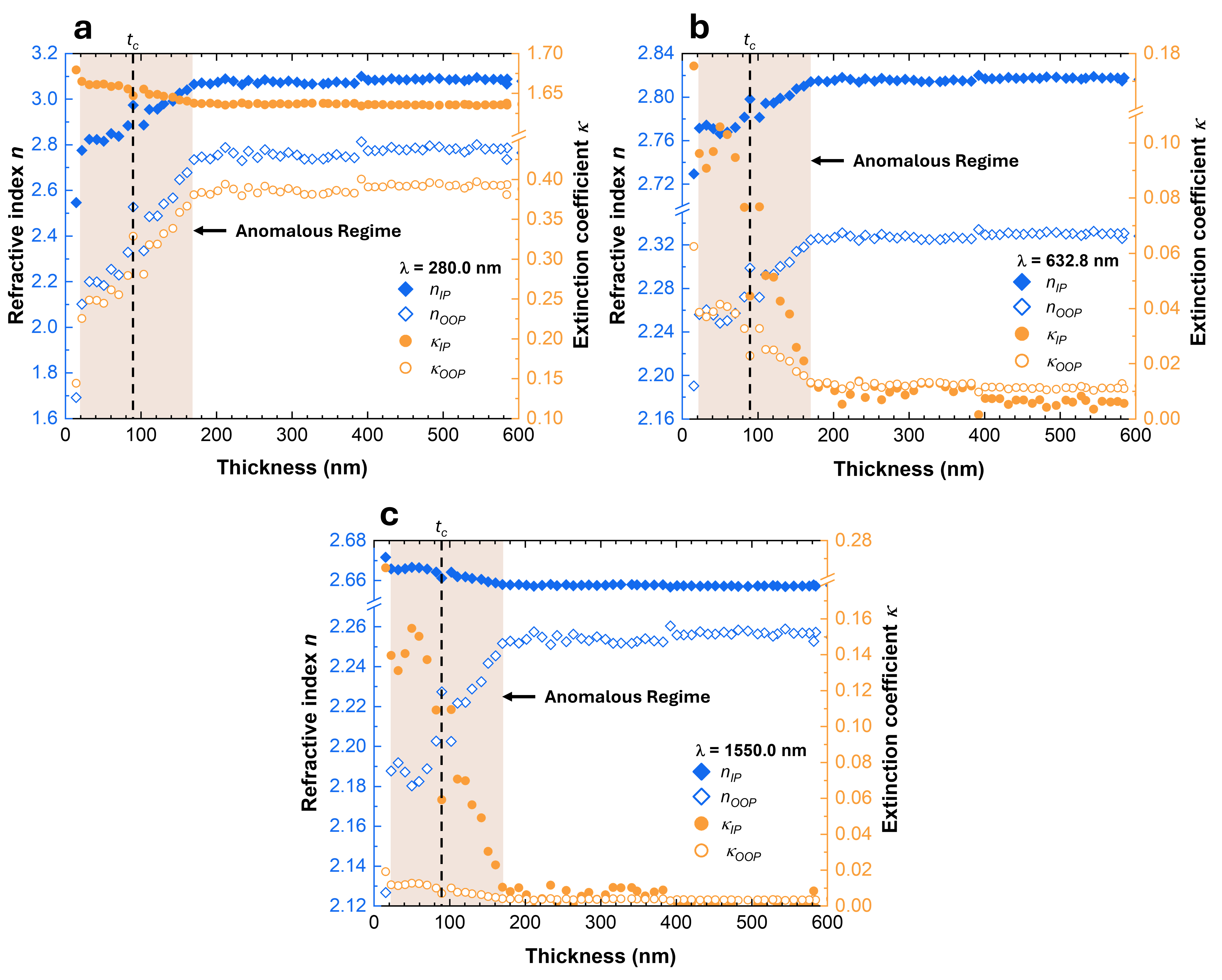}
\caption{\textbf{CIPS' optical constants thickness dependence.} In-plane and out-of-plane thickness dependence of CIPS' refractive index and extinction coefficient at three significant wavelengths: a) $\lambda =$ 280.0 nm, b) $\lambda =$ 632.8 nm, and c) $\lambda =$ 1550.0 nm. The orange shaded area corresponds to the anomalous thickness regime $t \in [22, 170)$ nm. The dashed line at $t \sim$ 90 nm corresponds to CIPS' thickness below which the IP polarizations disappears according to Ref. \cite{Deng2020}.}
\label{n vs thickness NEW}
\end{figure}


Figures \ref{n vs thickness NEW} and Figure \ref{dndt vs t} show the thickness-dependent IP and OOP refractive index and extinction coefficients of CIPS for $t \in$ (14, 600] nm at three significant wavelengths: $\lambda$ = 280.0 nm (the wavelength at which CIPS' change of refractive index with thickness is the largest, see Figure \ref{n,k vs lambda FIG2}c), $\lambda$ = 632.8 nm (commonly used wavelength in optics), and  $\lambda$ = 1550.0 nm (commonly used wavelength in telecommunications). For $t \geq 170$ nm, CIPS's optical constants remain constant with thickness as expected (albeit with an offset occurring at $t \sim$  390 nm). Within the thickness range $t \in$ [22, 170) nm, an anomalous thickness dependence is observed for both the IP and OOP optical constants, together with a large change in refractive index with thickness. Lastly, below $t \sim 22$ nm we encounter the thin limit, where finite size effects play a pivotal role in the material's properties and immense changes in CIPS' optical constants are observed, a phenomenon also reported for various 2D and 3D materials \cite{Wang2000, Hilfiker2017, Cai2010, Xu2013, Zhang2012, AitAli, Verrone2020}.
Considering that the electro-optic coefficient of bulk CIPS \cite{Liu2024} is comparable to typical 3D materials like \ce{LiNbO3} \cite{Hamze2020, jazbinsek}, CIPS flakes within the anomalous thickness regime are a promising platform for large electro-optic modulation. \\

For the range of wavelengths studied and within the anomalous regime, CIPS exhibits a thickness-dependent change in the OOP refractive as large as 23.2$\%$ at $\lambda = 280.0$ nm  (Table \ref{table1}). However, in the thickness range $t \in [22, 50]$ nm, CIPS' optical constants show a different trend compared to the rest of the anomalous regime. This change in habit could signify either a gradual leveling-off of the thickness-dependent effect or a change in mechanism for the behavior. Another sudden change is also observed for all optical constants, regardless of wavelength or direction, around $t \sim 90$ nm, coinciding with the critical thickness $t _c$  reported in Ref. \cite{Deng2020}, below which the IP polarization ceases to exist. Considering that the refractive index of a FE is coupled to its polarization \cite{Ross_2025, Haertling1987}, the rapid change in optical constants observed at $t \sim$ 90 nm may be explained by CIPS' IP polarization loss. Nevertheless, the anomalous behavior above and below $t$$_c$ cannot be explained by this behavior. To attempt to better understand the relationship between CIPS' crystal structure and its anomalous optical regime, we also carried out thickness-dependent Raman measurements. 

\subsection*{Thickness-dependent Raman spectroscopy}

\begin{figure}[H]
\centering
\includegraphics[width=1.00\textwidth, keepaspectratio]{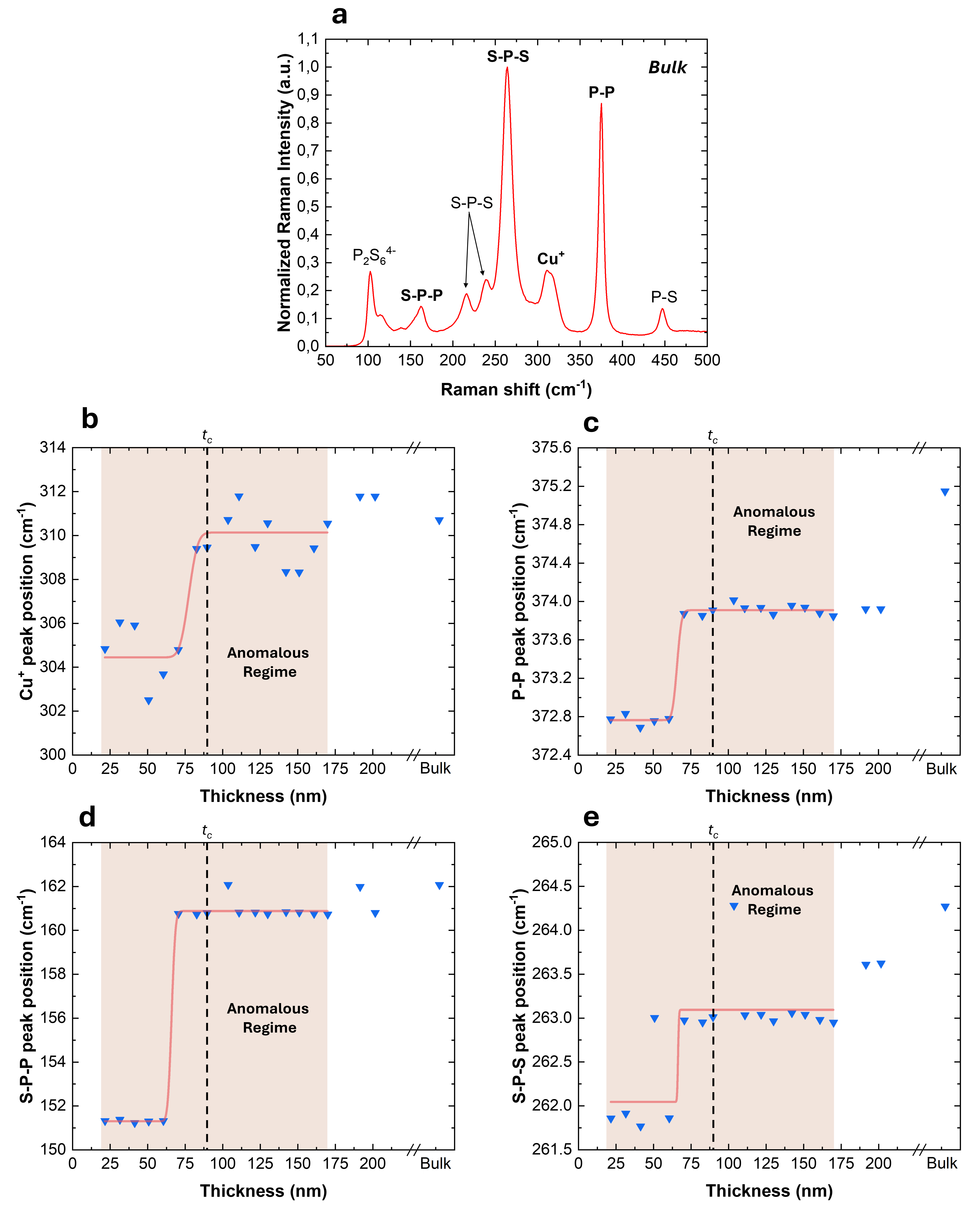}
\caption{\textbf{CIPS' Raman spectrum thickness dependence.} a) Bulk CIPS' normalized Raman spectrum in the range of Raman shifts $\omega \in$ [100, 400] cm$^{-1}$, with CIPS' vibrational modes ascribed to their respective peaks: $\omega$ $\sim$ 102.5 cm$^{-1}$, $\omega$ $\sim$ 114.5 cm$^{-1}$ $\rightarrow$ P$_2$S$_6^{-4}$, $\omega$ $\sim$ 162 cm$^{-1}$ $\rightarrow$ S-P-P, $\omega$ $\sim$ 216 cm$^{-1}$, $\omega$ $\sim$ 238.5 cm$^{-1}$, $\omega$ $\sim$ 264 cm$^{-1}$ $\rightarrow$ S-P-S, $\omega$ $\sim$ 311 cm$^{-1}$ $\rightarrow$ Cu$^+$, and $\omega$ $\sim$ 375 cm$^{-1}$ $\rightarrow$ P-P. In bold, the Raman shifts (vibrational modes) emphasized in this thickness-dependence study: b) $\omega(t) \sim 311$ cm$^{-1} \rightarrow$ \textbf{Cu$^+$}, c) $\omega$(t) $\sim$ 375 cm$^{-1}$ $\rightarrow$ \textbf{P-P}, d) $\omega$(t) $\sim$ 162 cm$^{-1}$ $\rightarrow$ \textbf{S-P-P}, and e) $\omega$(t) $\sim$ 264 cm$^{-1}$ $\rightarrow$ \textbf{S-P-S}. The shaded area corresponds to CIPS' anomalous thickness regime $t \in [22, 170)$ nm. The dashed line at $t \sim$ 90 nm corresponds to CIPS' thickness below which the IP polarizations disappears according to Ref. \cite{Deng2020}. The red lines correspond to data fitting using the error function.}
\label{raman peak position vs thickness}
\end{figure}

The Raman-active vibrational modes of CIPS were investigated between [100, 400] cm$^{-1}$ using Raman spectroscopy for thicknesses $t\in$ [22, 202] nm. Figure \ref{raman peak position vs thickness}a shows the normalized Raman spectrum of bulk CIPS, 
in good agreement with previous reports \cite{Thesis2023, Selhorst2024, PhysRevB.105.075151, Rao2022, Liu2024}. As seen in Figure \ref{FIG1}c, the absence of extra peaks as the CIPS flake is thinned from bulk to the few nanometer regime confirms that argon beam etching does not cause significant damage to CIPS' crystal structure \cite{ferroramantalk}. Furthermore, there are no signatures to indicate the formation of \ce{In$_{4/3}$P2S6} through Cu(I) deficiencies, which would lead to a different Raman fingerprint than observed for CIPS \cite{Rao2022}.\\

Figures \ref{raman peak position vs thickness}b-e present the thickness-dependent peak positions of four vibrational modes: 
Cu$^+$, S-P-P, S-P-S, and P-P. More peak positions are included in Figure \ref{raman peak position vs thickness extra}. Despite the presence of a few outliers, a clear shift in peak position is observed at $t\in$ [60, 71] nm for the P-P, S-P-S and S-P-P, while the shift for the Cu$^+$ peak can be found at $t\in$ [71, 83] nm. In both cases, the shift is sharp, distinctive of a first-order phase transition \cite{ferroramantalk}. For the case of the Cu$^+$ peak (Figure \ref{raman peak position vs thickness}b), the shift in peak position with thickness agrees reasonably well 
with the shift reported \cite{Xue2025}. \\




The shift in position of the P-P, S-P-P and S-P-S vibrational modes at $t\in$ [60, 71] nm might relate to CIPS' thickness-induced structural transition. In Figures \ref{n,k vs lambda FIG2}c-d, we observed shifts in the trends below $t\sim$ 60 nm, which may have been induced by this transition, potentially in parallel with other phenomena. Furthermore, the shift in position of the Cu$^+$ peak at $t\in$ [71, 83] nm may relate to changes in Cu(I) occupancy within its possible three positions. In Ref. \cite{Song2024} it was shown that the broad peak at $T = $ 300 K evolves into three peaks at $T = 78$ K. Analogously, we observed that while maintaining room-temperature conditions but decreasing thickness, the Cu$^+$ peak's width increases (Figure \ref{raman peak position vs thickness extra}a), implying an increase in Cu(I) disorder \cite{ferroramantalk} linked to a change in Cu(I) occupancies, hence a change in Cu(I) polarization.
Therefore, from the Raman data we believe that a change in CIPS' polarization takes place prior to its thickness-induced structural transition, which may explain the anomaly CIPS exhibits in the thickness-dependent optical constants at $t \sim$ 90 nm. \\



Having demonstrated that the CIPS' thickness-induced phase transformation is first-order and that changes in CIPS's polarization may already occur above its critical thickness $t_c\in$ [60, 71] nm, impacting its optical constants,
we proceed to compare the thickness-dependent refractive index of CIPS to commonly used dielectrics and FEs.

\subsection*{Comparison of thickness-dependent effects}
\begin{figure}[t]
\centering
\includegraphics[width=1.0\textwidth, keepaspectratio]{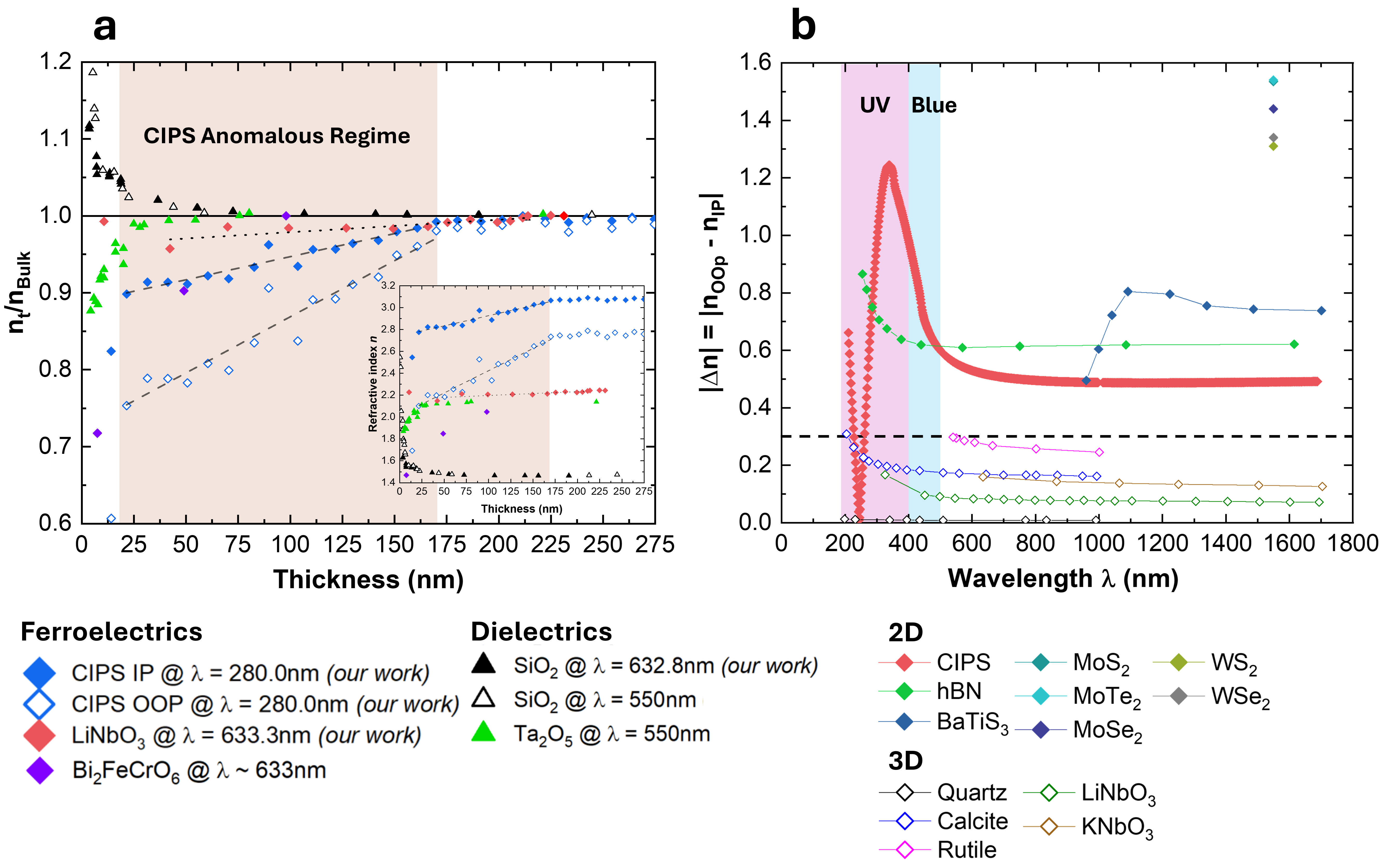}
\caption{\textbf{Optical properties for various materials.} a) Refractive index thickness dependence comparison between this work's materials (IP and OOP CIPS, \ce{LiNbO3} and \ce{SiO2}) and other materials in the literature (SiO$_2$ \cite{Cai2010}, Ta$_2$O$_5$ \cite{Zhang2012}, and \ce{Bi2FeCrO6} \cite{AitAli}). $n_t$/$n_{Bulk}$ represents the ratio between the refractive index at a certain thickness ($n_t$) and the refractive index at the maximum thickness ($n_{Bulk}$). The shaded area corresponds to CIPS' anomalous thickness regime $t \in$ [22, 170) nm. The dashed and the dotted lines show linear fittings of the CIPS and \ce{LiNbO3} data, respectively. The inset shows the raw refractive index thickness dependencies. b) Birefringence $\vert \Delta n \vert = \vert n_{OOP} - n_{IP} \vert$ comparison between CIPS at $ t \sim $ 22 nm and other materials in the literature: \ce{hBN} \cite{Grudinin2023}, \ce{BaTiS3} \cite{Niu2018}, \ce{MoS2} \cite{Ermolaev2021}, \ce{MoTe2}, \ce{MoSe2}, \ce{WS2}, \ce{WSe2} \cite{doi:10.1021/acsphotonics.2c00433}, quartz, calcite \cite{Ghosh1999}, rutile \cite{Sinton1961}, \ce{LiNbO3} \cite{Zelmon1997}, and \ce{KNbO3} \cite{Zysset1992}. The dashed line sets the threshold at which materials are defined to have giant birefringence. The violet and blue shaded areas correspond to the UV and blue wavelength ranges, respectively. }
\label{COMPARISONnew}
\end{figure}

To better understand the uniqueness of the anomalous thickness-dependent optical constants in CIPS, and as control experiments, we performed thickness-dependent VASE studies on two widely used materials, \ce{LiNbO3} and \ce{SiO2} (Figure \ref{LN n vs lambda and n vs t}). Figure \ref{COMPARISONnew}a shows the ratio between the refractive index at a certain thickness $n_t$ and the refractive index at the maximum thickness value $n_{Bulk}$ for different materials as a function of thickness. The refractive index of dielectric materials (e.g. SiO$_2$, Ta$_2$O$_5$) remains constant as thickness decreases ($n_t/n_{Bulk} \sim 1$) until the few-nanometer limit, where abrupt changes are expected and observed due to finite size effects. However, materials with ferroelectric nature, like 2D ferrielectric CIPS (this work), as well as 3D ferroelectric \ce{LiNbO3} (this work) and 3D multiferroic \ce{Bi2FeCrO6} \cite{AitAli}, show modulation of their refractive index as a function of thickness, 
 indicating that the anomalous behavior observed in CIPS may be generalizable to other ferroelectrics, but to a lesser extent. To the best of our knowledge, both CIPS's anomalous refractive index thickness dependence and its extension to other ferroelectrics have not been reported prior to this work, presenting a new tuning parameter for control. \\

 Figure \ref{COMPARISONnew}b shows CIPS's intrinsic birefringence $\vert \Delta n \vert = \vert n_{OOP} - n_{IP} \vert$ at $t \sim$ 22 nm as a function of wavelength compared to various contemporary birefringent materials. Notably, CIPS's birefringence, particularly in the blue-ultraviolet range, outperforms typically used 3D materials (quartz, calcite \cite{Ghosh1999, Smartt1959}, rutile \cite{Sinton1961}, and others \cite{Zelmon1997, Zysset1992}) as well as other low-dimension materials (2D: \ce{hBN} \cite{Grudinin2023}, transition metal dichalcogenides \cite{Ermolaev2021, doi:10.1021/acsphotonics.2c00433}, and others \cite{Guo2024, Yang2017, Mao2016}; quasi-1D: \ce{BaTiS3} \cite{Niu2018}; and 0D: \ce{[(p-C5H5NO)2ZnCl2]} \cite{znLi2024}), intrinsically, without the benefit of metasurface engineering \cite{Wang2024}. \\


\section*{Discussion}


To address the origin of the observed anomalous thickness-dependent change in CIPS' optical constants for $t \in$ (71, 170) nm, we briefly discuss five possible extrinsic and intrinsic mechanisms: substrate-induced strain, laser-induced heating, resonance between FE domain size and flake thickness, depolarization, and finally ferro-ionic effects. \\


Selhorst \textit{et al.} \cite{Selhorst2024} observed an increase in Raman peak frequencies for $t <$ 50 nm CIPS flakes, which they attribute to substrate-induced strain. 
If the origin of our observed Raman shifts is substrate-induced strain, as the thickness of the flake decreases, the strain suffered by the flake would increase \cite{Wu2022}, necessitating a continuous change of the Raman peak frequencies with thickness, similar to what Selhorst \textit{et al.} observe in the ultra-thin regime. Yet, we do not observe such behavior and given the lack of additional strain sources, 
substrate-induced effects appear to be insufficient to account for the observed trends, in agreement with the thickness-dependence Raman results reported in Ref. \cite{Xue2025}. \\

Another possible origin for the observed shifts in CIPS' vibrational modes could be 
laser-induced heating during Raman measurements. In the work from Dey \cite{Thesis2023}, non-negligible shifts in Raman peaks were observed using a laser power of 20 mW. Anticipating this issue, we used a much lower power of 0.5 mW. Additionally,
if our irradiation power was too large, we would expect the shift of the different vibrational modes to move to lower frequencies continuously with thickness, rather than the sudden step-like shift observed in Figures \ref{raman peak position vs thickness}b-e, making it unlikely for laser heating to account for our observations. \\

Beyond extrinsic mechanisms and regarding the thickness-dependent optical results of this work, we do not attribute the anomalous regime at $t \in$ [22, 170) nm 
to arise from FE domain formation of size similar to the thickness of the CIPS flake. 
In Ref. \cite{Chen2019} it was shown that CIPS' FE domains follow a Landau-Lifshitz-Kittel law within the thickness range $t \in$ (10, 130) nm, where $W \sim t^m$, with $W$ being the characteristic size of the FE domains, $t$ the thickness of the flake, and $m$ = 0.65, meaning that CIPS' FE domains would be much smaller than the thickness of the crystal.\\

Finally, as we thin the CIPS flake, we consider the depolarization field as a continuous source of polarization change with thickness \cite{depo, Li2009}, which would lead to continuous changes in the optical constants. 
However, the anomalous optical response observed in this work starts at a thickness $t \sim $ 170 nm; atypically high to be related to depolarization effects \cite{Gao2017}. Moreover, if depolarization was responsible for such an effect, we would expect 3D FE materials to show a more pronounced change in optical constants than 2D FE, as van der Waals FEs are more robust against depolarization \cite{Zhang2023}. Therefore, it is unlikely that depolarization fields explain the magnitude or the onset of CIPS's optical anomaly, motivating us to explore alternative intrinsic mechanisms. \\

Our work points towards a more compelling explanation, involving the interplay between light, ferroelectric polarization, and ionic mobility, an effect largely enhanced in the case of CIPS, compared to \ce{LiNbO3} and \ce{Bi2FeCrO6}, due to its distinct crystal structure. 
We suggest that the anomalous light-matter interaction may be influenced, or even mediated, by itinerant ionic charge carriers within the crystal structure. Ionic mobility is widely known within CIPS \cite{Maisonneuve1997, Maisonneuvle1997, Neumayer2025}, \ce{LiNbO3} \cite{Strelcov2014} and \ce{Bi2FeCrO6} \cite{Mahapatra2024}. However, this is especially significant in CIPS, known to be a ferro-ionic material, where both ferroelectric polarization and ionic conduction are intrinsically coupled through the Cu(I) spatial instability \cite{Maisonneuve1997, Maisonneuvle1997, Xu2020, Zhang2021}. Due to the strong electrostatic interaction between ferroelectric and ionic defect dipoles in CIPS, its polarization switching kinetics is known to be ionic-conduction-limited \cite{Zhou2020}, and when driven by an in-plane electric field, the movement of Cu(I) cations induces out-of-plane polarization switching \cite{Xu2020}. Moreover, in sister compound CuCrP$_2$S$_6$, it was recently shown that Cu(I) cation motion (via external field) can tune the refractive index \cite{Dushaq2024}. Thus, a change in CIPS' ion mobility while decreasing thickness could cause a change in the Cu(I) FE polarization contribution, resulting in a change in optical constants. Below $t_c \in [60,71]$ nm, the change in CIPS' ion mobility with thickness may be caused by its structural transition. At $t \in (71, 170)$ nm, the cause of a change of ion mobility with thickness remains elusive, but still with tremendous potential to be exploited for electro-optical and photonic applications.

\section*{Conclusion}
 We investigated the optical properties of the 2D ferrielectric \ce{CuInP2S6} over a wide range of thicknesses and wavelengths, and discovered a giant birefringence, maximized at $\vert \Delta n \vert$ $\sim$ 1.24 at $t \sim$ 22 nm and $\lambda$ = 339.5 nm, and the presence of an anomalous change in CIPS' refractive index with thickness in the range $t \in$ [22, 170) nm, as large as $\delta n \sim$ 23.2$\%$. To the best of our knowledge, this is the largest intrinsic birefringence $\vert n_{OOP} - n_{IP} \vert$ in the blue-ultraviolet regime of any known material, and the anomalous refractive index behavior reveals a new parameter to control optical response over a wide range of wavelengths and thicknesses, making CIPS an attractive system for electro-optic applications. This effect is shown to be the largest in CIPS, where both its polarization and ionic conductivity are intrinsically coupled; we ascribe changes in the optical response with thickness to changes in the Cu(I) polarization component. However, a complete explanation of the entire anomalous thickness regime is yet to be determined. We propose a complex and strong thickness-dependent interplay between polarization, light and ion mobility as an explanation for the anomalous nature of CIPS with thickness that we show may be extendable to other ferroelectrics, like \ce{LiNbO3}, or multiferroics, like \ce{Bi2FeCrO6}, though to a lesser extent compared to CIPS. Further investigation is needed to elucidate such complex interactions, yet our work paves the way for a paradigm shift in electro-optics and photonics based on CIPS' family of materials.  \\

\textbf{Acknowledgments: }
The authors acknowledge Heng Wu, Yaojia Wang, Michiel Dubbelman and Victor Vreede for their valuable discussions. We also thank Lodi Schriek (spectroscopic ellipsometry and Raman spectroscopy) and Charles de Boer (argon beam etcher) for their experimental support. The authors thank Yanyu Liu and Jiawang Hong for providing the CIPS' trigonal \textit{P}3$_1$\textit{c} CIF file for the preparation of Figure \ref{FIG1}a bottom panel.\\

\textbf{Funding:} The authors acknowledge the support from the Dutch Research Council (NWO) under the 2023 VIDI Award, as well as the Kavli Foundation under the Kavli Institute Innovation Award (KIIA).\\

\textbf{Author contributions:} H. M. H. conceived the study. H. M. H. and R. J. H. K. designed the study.  R. J. H. K. conceived the low-angle argon beam etching. H. M. H. performed the argon beam etching and the spectroscopic ellipsometry and Raman spectroscopy measurements.  R. J. H. K. created the spectroscopic ellipsometry model. H. M. H. carried out the data analysis. H. M. H. wrote the majority of the manuscript. T. M. K, R. J. H. K., and M. N. A. reviewed the manuscript and provided theoretical support and discussion. M. N. A. is the principal investigator. All authors contributed to the preparation of the manuscript.\\

\textbf{Competing interests:} The authors declare that they have no competing interests.\\

\textbf{Data and materials availability:} The data that supports the findings of this study are available from 
the corresponding authors upon reasonable request.\\

\renewcommand{\thefigure}{S\arabic{figure}}

\setcounter{figure}{0}

\section*{Methods}
\label{sectionmethods}

\subsection*{Sample preparation}
The CIPS flake used in this work was mechanically exfoliated directly from a bulk CIPS single crystal (HQ graphene) onto a Si substrate with a thermally grown \ce{SiO2} layer $\sim$ 300 nm thick (see Figure \ref{FIG1}e top left microscope image). After exfoliation, spectroscopic ellipsometry and Raman spectroscopy measurements were performed.


\subsection*{Spectroscopic ellipsometry} 
\subsubsection*{Measurement}
The optical constants and thickness of the CIPS sample were measured at room temperature using a Woollam M2000XI-210 Variable Angle Spectroscopic Ellipsometer with a spectral range between 210.6 nm and 1688.3 nm (see Figure \ref{FIG1}b). Before starting each of the actual measurements, a mapping scan of the whole flake was performed with a step size of 4.67 $\mu$m to select the cleanest and least noisy regions of the CIPS flake plateau. The actual measurements were performed at three different angles of incidence: 65, 70 and 75 degrees. Focus probes were used to reduce the beam spot size to $\sim$ 300 $\mu$m at 70 degrees. During the measurement, due to the presence of a rotating compensator before the sample and a rotating polarizer and analyzer in the ellipsometer, all elements of the Mueller matrix with the exception of the bottom row (M41 to M44) were measured. \\ 

\subsubsection*{Modelling} 

The ellipsometry data was analysied using  the CompleteEase software. An example of the raw ellipsometric spectrum and the corresponding fitting is shown in Figure \ref{etch rate}a. The main model is split up into two submodels, a bulk and a thin regime model, and uses a Bruggeman effective medium approximation (EMA) between them. Both the bulk and thin regime models were established using a multi-sample analysis, which enabled simultaneous fitting of different measurements. In this way,  the quality of the model increases, as it needs to fit more varied data. Both models start off with a Cauchy-Urbach model to explain the dispersive regions of the spectrum. The Cauchy-Urbach is subsequently fed into a basis spline function with a node distance of 0.5 eV. To make this raw mathematical model physical, a combination of a Cody-Lorentz oscillator and a Gaussian oscillator was built. The same process was followed for the in-plane and out-of-plane  axes. Finally, the thin and bulk models were combined with the Bruggeman EMA model. Note that both Maxwell-Garnet and Bruggeman approaches yielded nearly identical relation between thickness and refractive index. Considering the properties of CIPS, the most appropriate choice was chosen to be the Bruggeman approximation.\\

Besides the main model, certain imperfections were also taken into account.  A roughness model (consisting of a Bruggeman EMA of 50$\%$ air and 50$\%$ the current model), a thickness non-uniformity depolarization fit and bandwidth error of the two spectrum analyzers (using a Gaussian overlap function between the 5 nm bandwidth UV-VIS and 10 nm bandwidth IR spectrum analyzers) were used.\\

Other models were also tested as a part of the investigation. However, the most obvious other options (such as graded models) did not yield better fitting results. Therefore, it implies that, in fact, CIPS' optical properties actually change with thickness. It should also be noted that the mean square error between fit and model (MSE) was best minimized using anisotropic models and showed improvement in MSE in all cases tested, as expected with van der Waals materials 
\cite{doi:10.1021/acsphotonics.2c00433}.\\

Lastly, it has been reported that CIPS also exhibits a small but non-negligible in-plane optical anisotropy $\Delta n = n_x -n_y = 0.0149$ \cite{Liu2024}. Nevertheless, including trirefringence instead of birefringence did not considerably improve the model, as it is less sensitive to in-plane birefringence. Thus, in-plane optical anisotropy was not considered in this work. \\

\subsection*{Raman spectroscopy}
\subsubsection*{Measurement}
For the bulk flake and the last 20 etches ( $t \in$ [22, 202] nm), Raman spectroscopy measurements were included in the measurement cycle, as shown in Figure \ref{FIG1}c. A Renishaw Invia Reflex Raman microscope was used at room temperature. Before every measurement, a Si reference substrate was measured to calibrate the instrument. Each spectrum was obtained  with an excitation wavelength of 488 nm, a grating of 3000 lines/mm, a power of 0.50 mW, 10 accumulations, 10 s acquisition time, and a 50x objective. The spot size of the Raman microscope used is 1.5 $\mu$m. For each thickness, a Raman line scan with a step size of 1 $\mu$m was performed across the same homogeneous region of the CIPS flake, obtaining  $\geq$10 Raman spectra respectively.

\subsubsection*{Data analysis}

The $\geq$10 spectra were averaged and normalized to the $\sim$ 375 cm$^{-1}$ peak (most intense peak). Peaks were found using OriginPRO 2024 Find Peaks tool. First, we set a user defined baseline using 8 anchor points. Then, the baseline was fitted using a linear function (an exponential decay fitted baseline was also tried but results did not change significantly). Finally, peaks were found via local maximum method, using 2-4 local points and a peak filtering by height from 2-10$\%$, depending on the noisiness of the data.\\ 




\subsection*{Argon beam etching}
Once the CIPS flake was characterized via ellipsometric and Raman spectroscopies, the sample was dry-etched under a 4 degree sheering angle and a 10 rpm stage rotation with a SCIA Mill150 argon beam etcher, using a beam voltage of 200 V and an acceleration beam voltage of 100 V, as illustrated in Figure \ref{FIG1}d. 
In this etcher, the accelerated Ar ions are neutralized before hitting the sample using electrons produced by a filament under vacuum, with a filament current of 250 mA. All the parameters of the argon beam etcher were not changed during the entirety of this work, independently of the material etched  (CIPS, \ce{LiNbO3} and \ce{SiO2}) except for the etching time, obtaining a constant etch rate, as shown in Figure \ref{etch rate}b for the CIPS' etching. 
\section*{Extended data}

\begin{figure}[H]
\centering
\includegraphics[width=1.0\textwidth, keepaspectratio]{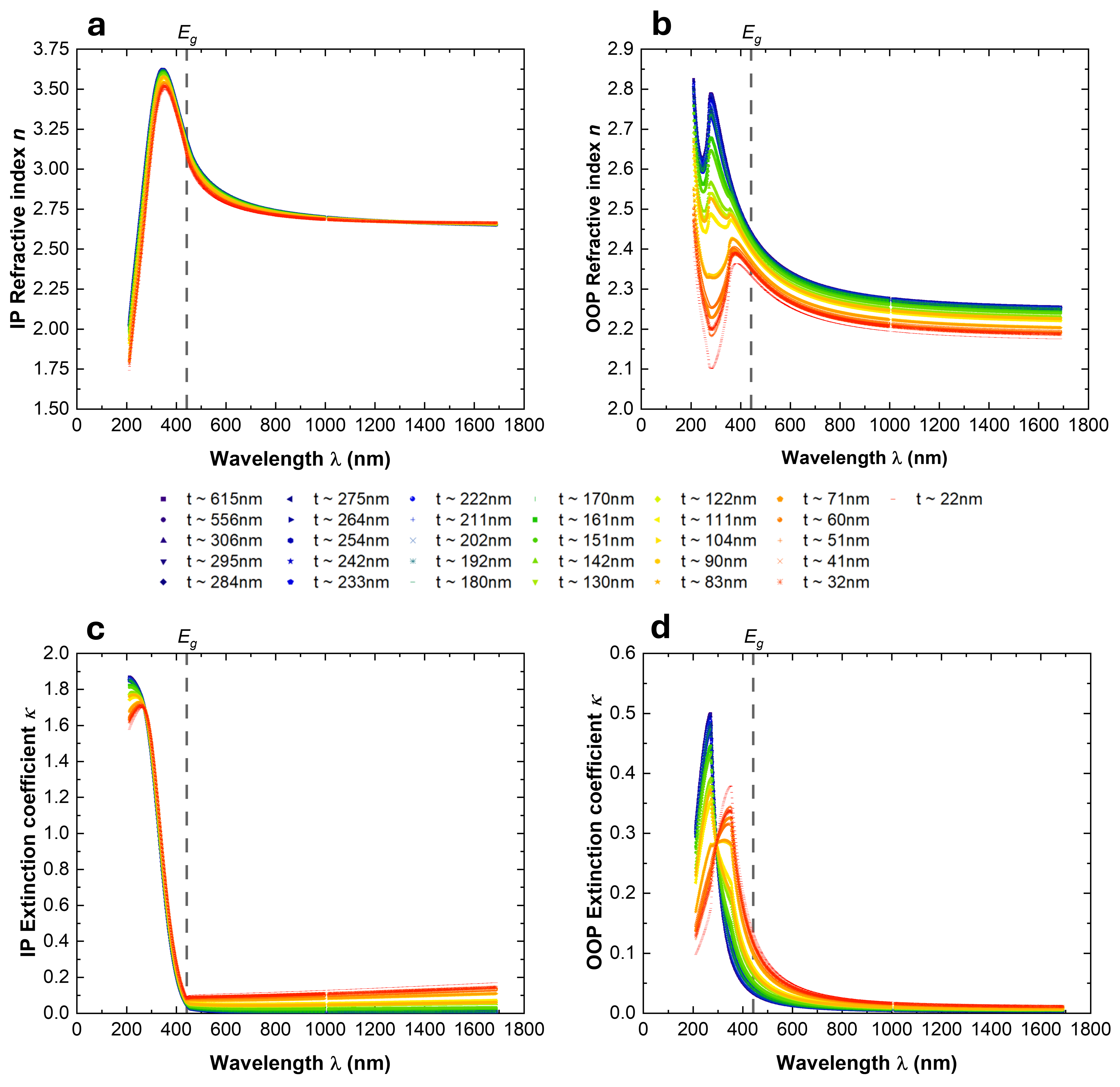}
\caption{\textbf{Full CIPS' optical constants wavelength dependence.} a) In-plane and b) Out-of-plane CIPS' refractive index wavelength dependence for the whole range of thicknesses studied $t \in [22, 615)$ nm. c) In-plane and d) Out-of-plane CIPS' extinction coefficient wavelength dependence for the whole range of thicknesses studied $t \in [22, 615)$ nm. The dashed lines correspond to the reported CIPS' band gap value $E_g \sim$ 2.8 eV ($\lambda \sim$  443 nm) \cite{Zhou2021, Ho2021, Bu2023}.}
\label{Full n vs lambda}
\end{figure}

\begin{figure}[H]
\centering
\includegraphics[width=0.6\textwidth, keepaspectratio]{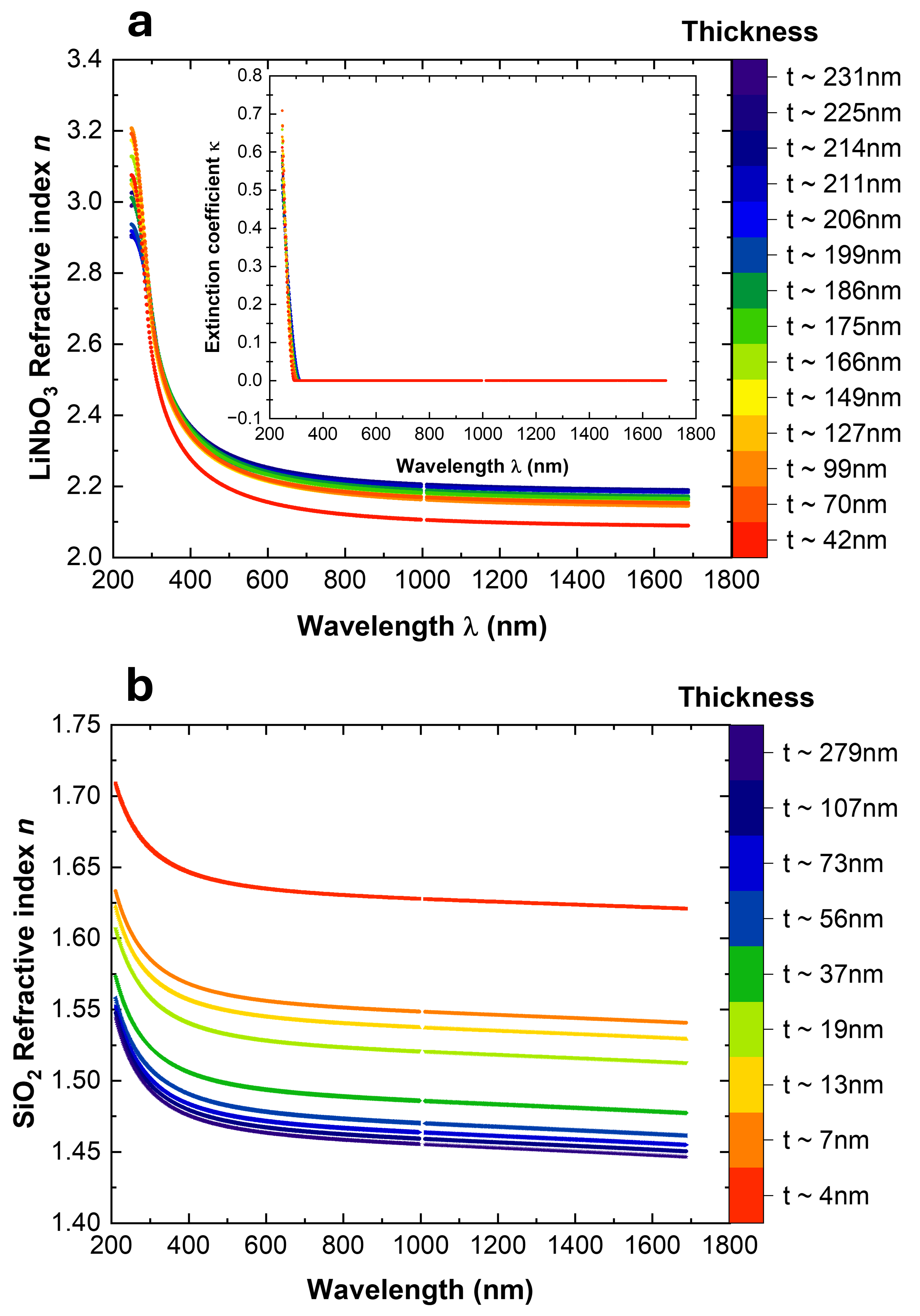}
\caption{ \textbf{Reference materials' optical constants wavelength dependence.} a)\ce{LiNbO3} refractive index wavelength dependence for all \ce{LiNbO3} thicknesses measured in this work. The inset shows the \ce{LiNbO3} extinction coefficient wavelength dependence. b) \ce{SiO2}  refractive index wavelength dependence for all SiO$_2$ thicknesses measured in this work.}
\label{LN n vs lambda and n vs t}
\end{figure}

\begin{figure}[H]
\centering
\includegraphics[width=1.0\textwidth, keepaspectratio]{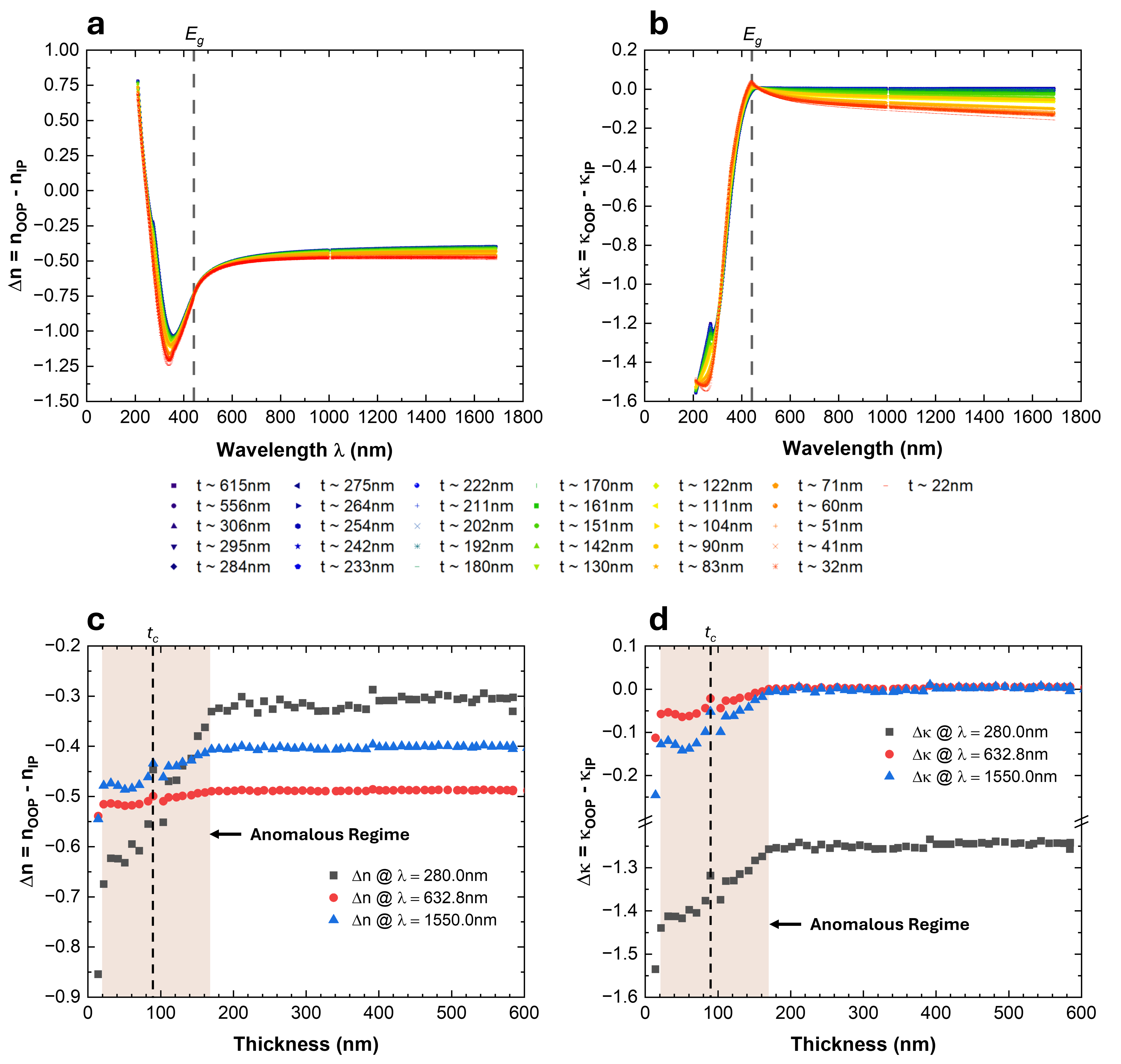}
\caption{ \textbf{CIPS' anisotropy.} Wavelength dependence at different thicknesses of CIPS' a) Birefringence $\Delta n = n_{OOP} - n_{IP}$ and b) Dichroism $\Delta \kappa = \kappa_{OOP} - \kappa_{IP}$ for the whole range of thicknesses studied $t \in [22, 615)$ nm. The dashed lines correspond to the reported CIPS' band gap value $E_g \sim$ 2.8 eV ($\lambda \sim$  443 nm) \cite{Zhou2021, Ho2021, Bu2023}. Thickness dependence of CIPS' c) Birefringence and d) Dichroism at wavelengths $\lambda$ = 280.0 nm, 632.8 nm and 1550.0 nm. The dashed line at $t \sim$ 90 nm corresponds to CIPS' thickness below which the IP polarizations disappears according to Ref. \cite{Deng2020}.}
\label{Anisotropy vs thickness vs lambda}
\end{figure}

\begin{figure}[H]
\centering
\includegraphics[width=0.98\textwidth, keepaspectratio]{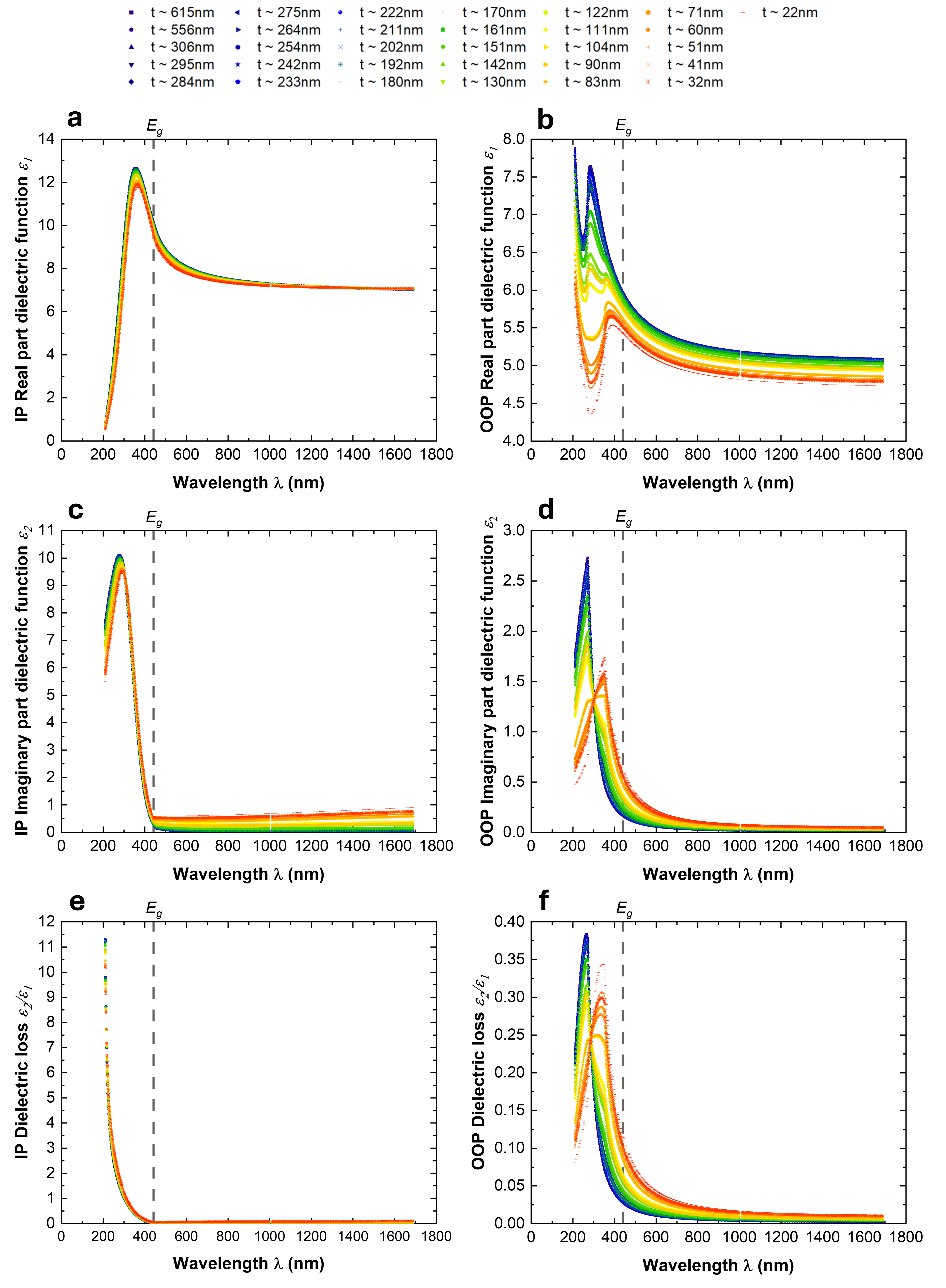}
\caption{\textbf{Full CIPS' dielectric function wavelength dependence.} a) In-plane and b) Out-of-plane CIPS' real part of the dielectric function ($\varepsilon_1$) wavelength dependence for the whole range of thicknesses studied $t \in [22, 615)$ nm. c) In-plane and d) Out-of-plane CIPS' imaginary part of the dielectric function ($\varepsilon_2$) wavelength dependence for the whole range of thicknesses studied $t \in [22, 615)$ nm. e) In-plane and f) Out-of-plane CIPS' dielectric loss ($\varepsilon_2/\varepsilon_1$) wavelength dependence for the whole range of thicknesses studied $t \in [22, 615)$ nm. The dashed lines correspond to the reported CIPS' band gap value $E_g \sim$ 2.8 eV ($\lambda \sim$  443 nm) \cite{Zhou2021, Ho2021, Bu2023}.}
\label{dielectric function vs lambda}
\end{figure}

\begin{figure}[H]
\centering
\includegraphics[width=1.0\textwidth, keepaspectratio]{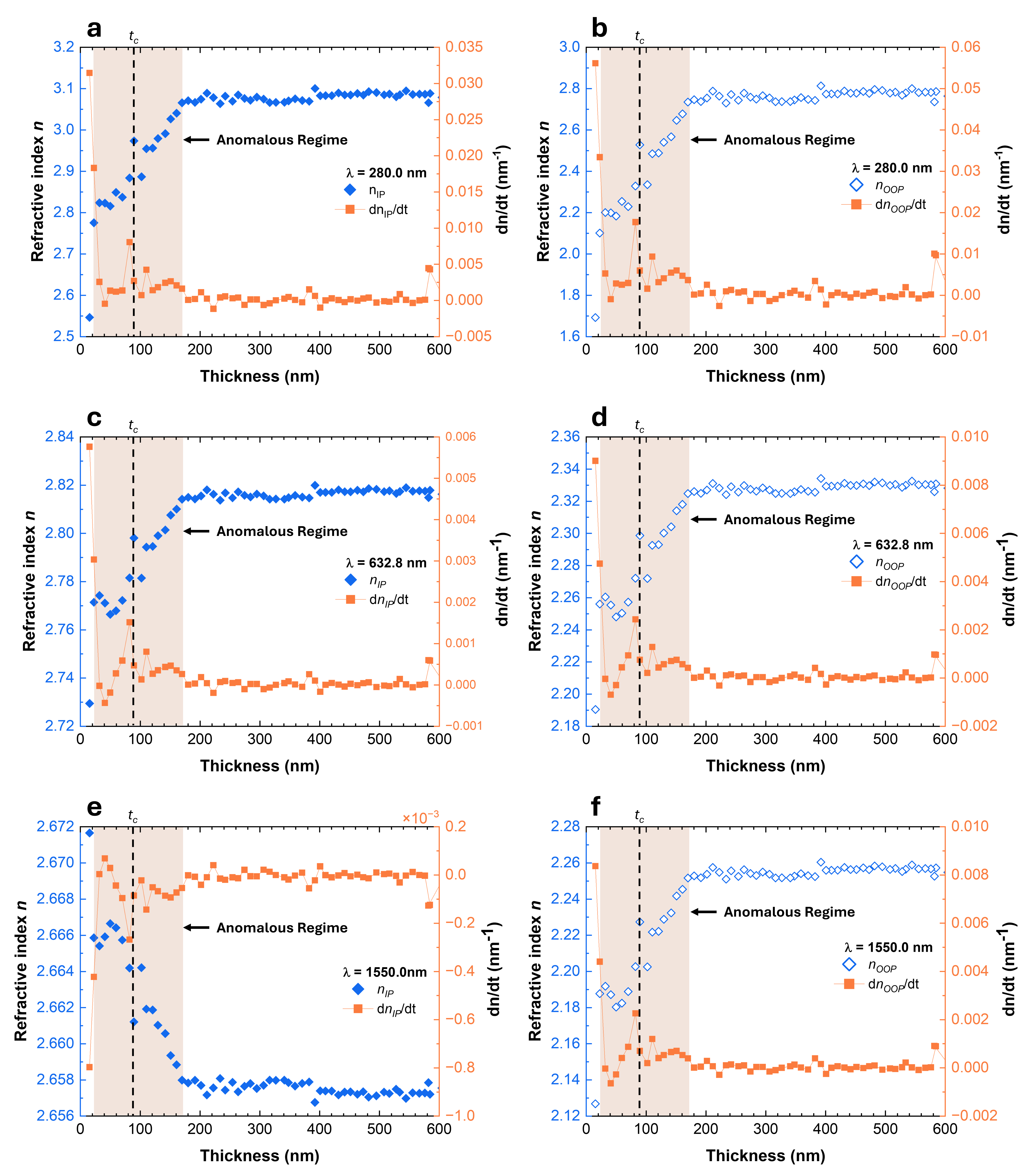}
\caption{\textbf{ CIPS' refractive index derivative thickness dependence.} a) In-plane and b) out-of-plane CIPS' refractive index ($n_{IP}$) and derivative of the in-plane refractive index with respect to thickness (d$n_{IP}$/dt) versus thickness at wavelength $\lambda$ = 280.0 nm.  c) In-plane and d) out-of-plane refractive index ($n_{IP}$) and derivative of the in-plane refractive index with respect to thickness (d$n_{IP}$/dt) versus thickness at wavelength $\lambda$ = 632.8 nm.  e) In-plane and f) out-of-plane refractive index ($n_{IP}$) and derivative of the in-plane refractive index with respect to thickness (d$n_{IP}$/dt) versus thickness at wavelength $\lambda$ = 1550.0 nm.  The shaded area corresponds to the anomalous thickness regime $t \in [22, 170)$ nm. The dashed line at $t \sim$ 90 nm corresponds to CIPS' thickness below which the IP polarizations disappears according to Ref. \cite{Deng2020}.}
\label{dndt vs t}
\end{figure}

\begin{figure}[H]
\centering
\includegraphics[width=1.0\textwidth, keepaspectratio]{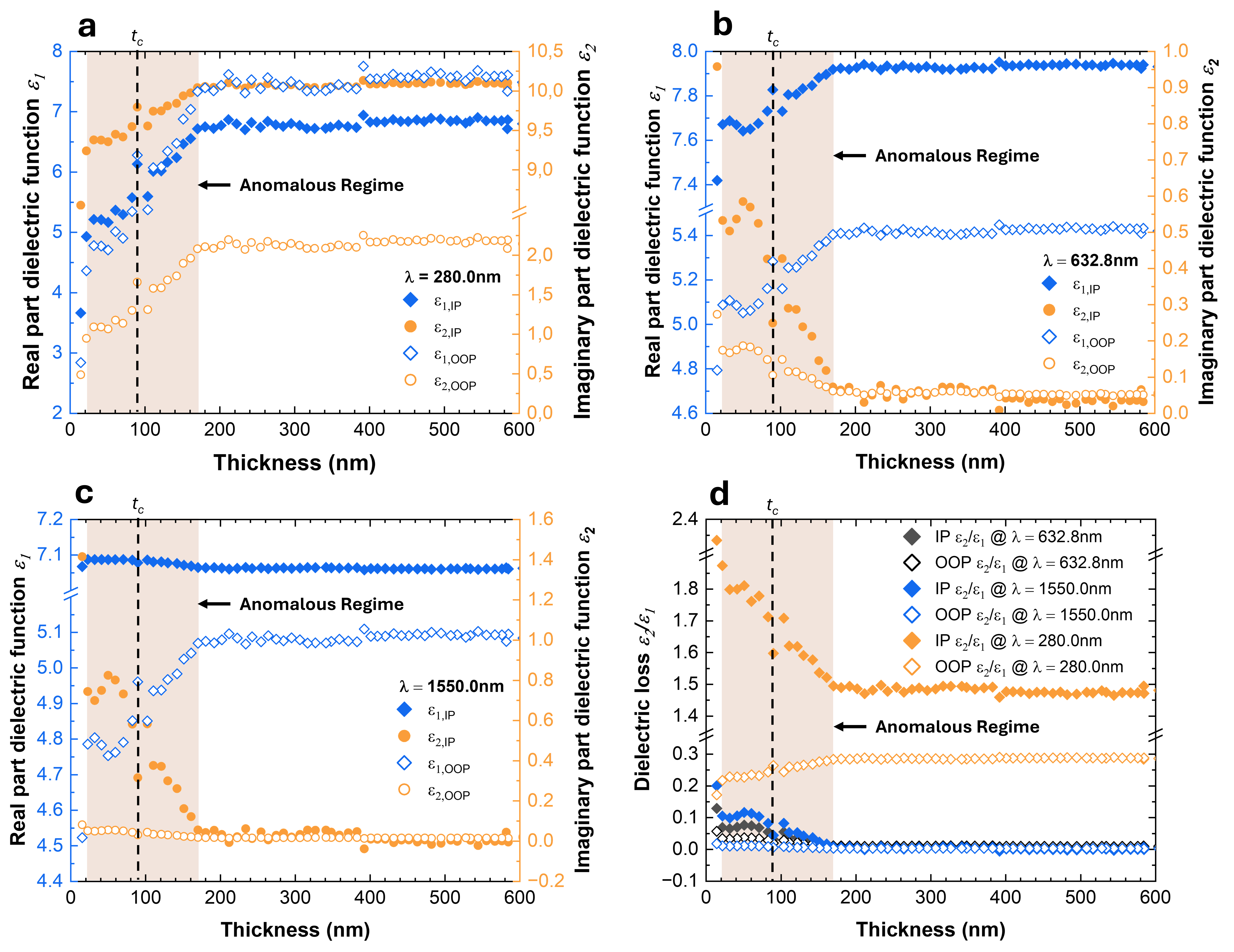}
\caption{\textbf{CIPS' dielectric function thickness dependence.} In-plane and out-of-plane thickness dependence of CIPS' dielectric function ($\hat{\varepsilon} = \varepsilon_1 + i\varepsilon_2$)) at different wavelengths: a) $\lambda =$ 280.0 nm, b) $\lambda =$ 632.8 nm, c) $\lambda =$ 1550.0 nm. d) In-plane and out-of-plane thickness dependence of CIPS' dielectric loss ($\varepsilon_2/\varepsilon_1$) at three wavelengths: $\lambda =$ 280.0 nm, $\lambda =$ 632.8 nm, $\lambda =$ 1550.0 nm. The shaded area corresponds to the anomalous thickness regime $t \in [22, 170)$ nm. The dashed line at $t \sim$ 90 nm corresponds to CIPS' thickness below which the IP polarizations disappears according to Ref. \cite{Deng2020}.}
\label{dielectric function vs thickness}
\end{figure}

\begin{figure}[H]
\centering
\includegraphics[width=1.0\textwidth, keepaspectratio]{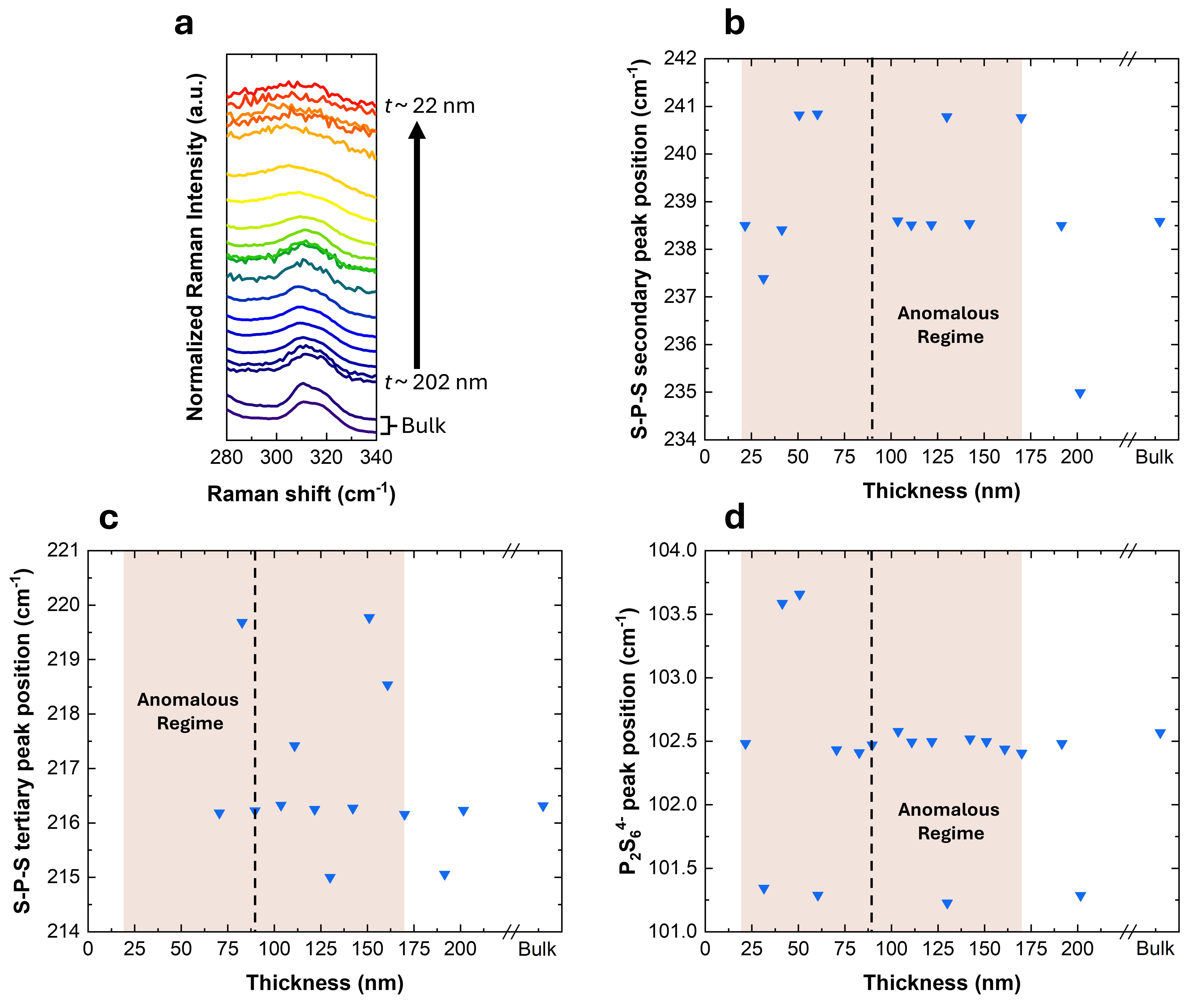}
\caption{\textbf{Extended CIPS' Raman spectra thickness dependence.} a) Collection of all normalized CIPS' Raman spectra from bulk to $t \sim$ 22 nm zoomed-in around the Cu$^+$ peak position. Raman shifts with non-conclusive thickness dependence trend:  b) $\omega(t) \sim 238.5$ cm$^{-1} \rightarrow$ S-P-S, c) $\omega(t) \sim 216$ cm$^{-1} \rightarrow$ S-P-S (for $t<70$nm, the amplitude of the peak was at noise level and could not be determined), and d) $\omega(t) \sim 102.5$ cm$^{-1} \rightarrow$ P$_2$S$_6^{-4}$.  Due to the noise in the raw Raman data, the values of such peak positions have a reduced confidence compared to the peaks showed in Figure \ref{raman peak position vs thickness}, with some peaks being undetectable at certain thicknesses (without increasing the laser power, which would increase the risk of burning the sample). The shaded area corresponds to CIPS' anomalous thickness regime $t \in [22, 170)$ nm. The dashed line at $t \sim$ 90 nm corresponds to CIPS' thickness below which the IP polarizations disappears according to Ref. \cite{Deng2020}. }
\label{raman peak position vs thickness extra}
\end{figure}

\begin{figure}[H]
\centering
\includegraphics[width=1.0\textwidth, keepaspectratio]{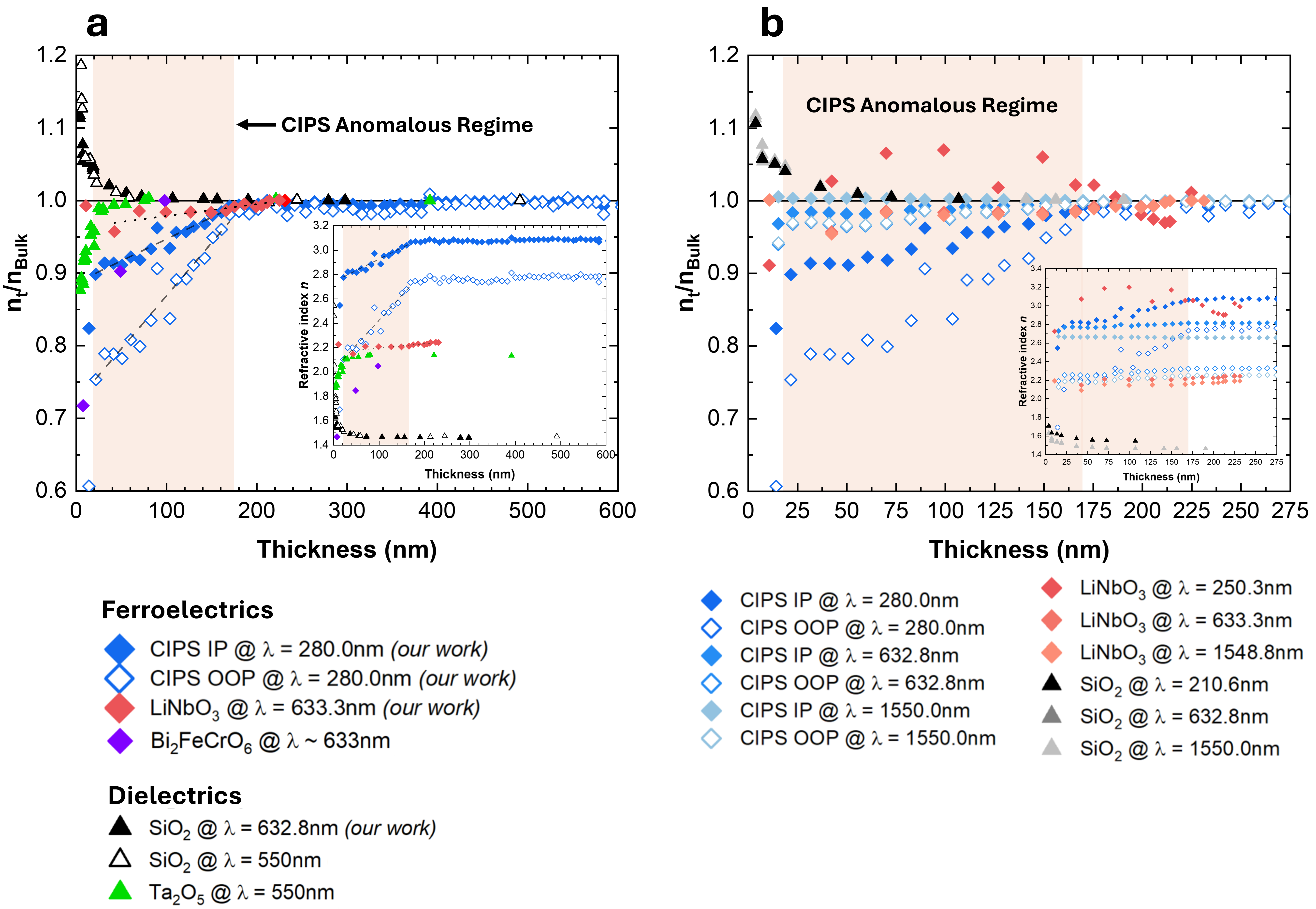}
\caption{\textbf{Extended refractive index thickness dependence for various materials.} a) Comparison between this work's measured materials (IP and OOP CIPS, \ce{LiNbO3} and \ce{SiO2}) and results in the literature (SiO$_2$ \cite{Cai2010}, \ce{Ta2O5} \cite{Zhang2012}, and \ce{Bi2FeCrO6} \cite{AitAli}. $n_t$/$n_{Bulk}$ represents the ratio between the refractive index at a certain thickness ($n_t$) and the refractive index at the maximum thickness ($n_{Bulk}$). The shaded area corresponds to CIPS' anomalous thickness regime $t \in [22, 170)$ nm. The dashed and the dotted lines show linear fittings of the CIPS and \ce{LiNbO3} data, respectively. The inset shows the raw refractive index thickness dependencies. b) $n_t$/$n_{Bulk}$ thickness-dependence comparison of the three materials studied in this work (CIPS, \ce{LiNbO3} and SiO$_2$) at different wavelengths respectively.The inset shows the raw refractive index thickness dependencies. }
\label{COMPARISONFULL}
\end{figure}

\begin{figure}[H]
\centering
\includegraphics[width=1.0\textwidth, keepaspectratio]{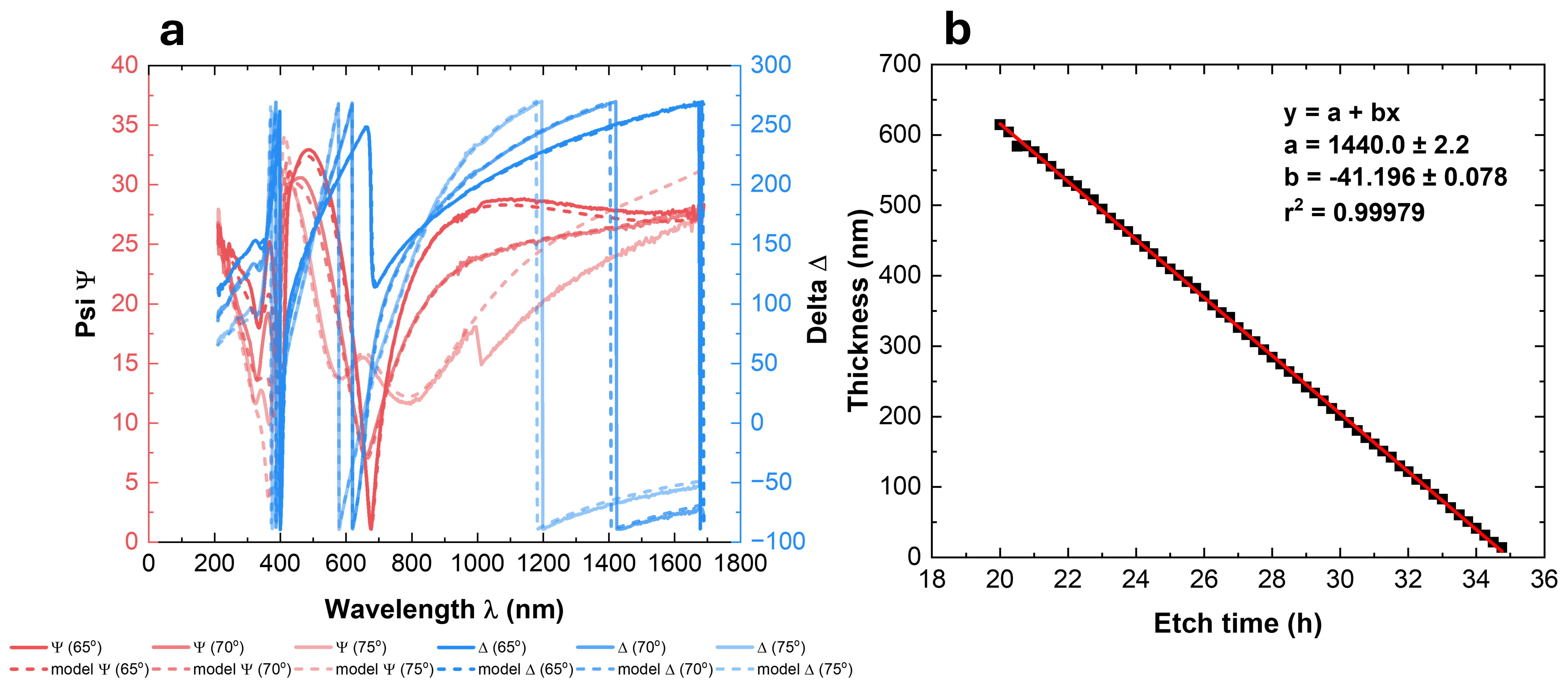}
\caption{\textbf{CIPS' characterization.} a) Ellipsometric spectrum taken at three different incident beam angles: 65$^{\circ}$, 70$^{\circ}$ and 75$^{\circ}$. $\Psi$ and $\Delta$ relate to the change in light polarization when the beam interacts with the surface of the sample, such that: $\tan{(\Psi}) e^{i\Delta} = r_p/r_s$, where $r_p$ and $r_s$ are the Fresnel reflection coefficients for the parallel and perpendicularly polarized light respectively. From this spectrum, a model can be built as described in the \nameref{sectionmethods}. The results of the model are shown as dotted lines. The extracted CIPS' thickness from the model is $t \sim$ 60 nm. b) Change of CIPS' thickness with etch time at a sheering angle of 4 $^{\circ}$, under a beam voltage of 200 V, an accelaration beam voltage of 100 V, and a filament current of 250 mA. CIPS' etch rate at such conditions is $\Delta t \sim$ 41 nm/h.}
\label{etch rate}
\end{figure}

\renewcommand{\thetable}{S\arabic{table}}
\renewcommand{\thefootnote}{\fnsymbol{footnote}}
\begin{table}[b]
\centering
\caption{\textbf{Change in IP and OOP refractive index ($\delta n$) extracted from ellipsometry for different wavelengths and thickness regimes.}
}
\label{table1}%
\begin{tabular}{@{}cccc@{}}
\toprule
Wavelength (nm) & Axis direction & Thickness regime (nm) & $\delta n$\footnotemark[1] ($\%$) \\
\midrule
\multirow{4}{*}{280.0}   & IP   & $585\geq t\geq170$  & 0.73 \\
   & OOP   & $585\geq t\geq170$  & 1.84  \\
    & IP   & $22 \leq t<170$  & 9.47 \\
   & OOP   & $22 \leq t<170$  & \textbf{23.18}  \\
   \midrule
\multirow{4}{*}{632.8}   & IP   & $585\geq t\geq170$  & 0.13 \\
   & OOP   & $585\geq t\geq170$  & 0.26  \\
   & IP   & $22 \leq t<170$  & 1.52 \\
   & OOP   & $22 \leq t<170$  & 2.96 \\
\midrule
\multirow{4}{*}{1550.0}   & IP  & $585\geq t\geq170$  & -0.03  \\
   & OOP  & $585\geq t\geq170$  & 0.25  \\
    & IP   & $22 \leq t<170$  & -0.30 \\
   & OOP   & $22 \leq t<170$  & 2.84 \\
\hline

\botrule
\end{tabular}
\footnotemark[1]{$\delta n = \frac{n_{max}-n_{min}}{n_{max}}\cdot 100$, where $n_{max}$ is the refractive index at the maximum value of the thickness regime and $n_{min}$ is the refractive index at the minimum value of the thickness regime.}\\
\end{table}

\bibliography{sn-bibliography}

\end{document}